L. Bouchaoui, K.E. Hemsas, H. Mellah, S. Benlahneche

# POWER TRANSFORMER FAULTS DIAGNOSIS USING UNDESTRUCTIVE METHODS (ROGER AND IEC) AND ARTIFICIAL NEURAL NETWORK FOR DISSOLVED GAS ANALYSIS APPLIED ON THE FUNCTIONAL TRANSFORMER IN THE ALGERIAN NORTH-EASTERN: A COMPARATIVE STUDY

*Introduction. Nowadays, power transformer aging and failures are viewed with great attention in power transmission industry. Dissolved gas analysis (DGA) is classified among the biggest widely used methods used within the context of asset management policy to detect the incipient faults in their earlier stage in power transformers. Up to now, several procedures have been employed for the lecture of DGA results. Among these useful means, we find Key Gases, Rogers Ratios, IEC Ratios, the historical technique less used today Doernenburg Ratios, the two types of Duval Pentagons methods, several versions of the Duval Triangles method and Logarithmic Nomograph. Problem. DGA data extracted from different units in service served to verify the ability and reliability of these methods in assessing the state of health of the power transformer. Aim. An improving the quality of diagnostics of electrical power transformer by artificial neural network tools based on two conventional methods in the case of a functional power transformer at Sétif province in East North of Algeria. Methodology. Design an inelegant tool for power transformer diagnosis using neural networks based on traditional methods IEC and Rogers, which allows to early detection faults, to increase the reliability, of the entire electrical energy system from transport to consumers and improve a continuity and quality of service. Results. The solution of the problem was carried out by using feed-forward back-propagation neural networks implemented in MATLAB- Simulink environment. Four real power transformers working under different environment and climate conditions such as: desert, humid, cold were taken into account. The practical results of the diagnosis of these power transformers by the DGA are presented. Practical value. The structure and specific features of power transformer winding insulation ageing and defect state diagnosis by the application of the artificial neural network (ANN) has been briefly given. MATLAB programs were then developed to automate the evaluation of each method. This paper presents another tool to review the results obtained by the delta X software widely used by the electricity company in Algeria.* References 29, table 15, figures 9.

*Вступ. У наш час старіння та несправності силових трансформаторів уважно розглядаються у галузі передачі електричної енергії. Аналіз розчиненого газу виділяється серед найбільш широко використовуваних методів, що застосовуються в контексті політики управління активами для виявлення початкових несправностей на їх попередній стадії в силових трансформаторах. Дотепер для отримання результатів аналізу розчиненого газу було використано кілька процедур. Серед цих корисних засобів зазначимо такі, як метод основних газів, коефіцієнти Роджерса, коефіцієнти МЕК, історичний підхід, менш використовувані сьогодні коефіцієнти Дерненбурга, два типи методів п'ятикутників Дюваля, кілька варіантів методу трикутників Дюваля та логарифмічний номограф. Проблема. Дані аналізу розчиненого газу, отримані з різних об'єктів, що експлуатуються, слугували для перевірки здатності та надійності цих методів при оцінці стану працездатності силового трансформатора. Мета. Підвищення якості діагностики електричного силового трансформатора за допомогою штучних нейромережевих інструментів, заснованих на двох звичайних методах, у випадку функціонуючого силового трансформатора в провінції Сетіф на північному сході Алжиру. Методологія. Розробка нетипового засобу для діагностики силових трансформаторів з використанням нейронних мереж на основі традиційних методів МЕК і Роджерса, який дозволяє раннє виявлення несправностей, підвищення надійності всієї електроенергетичної системи від передачі енергії до споживачів та покращення безперервності та якості обслуговування. Результати. Розв'язання проблеми було здійснено за допомогою нейронних мереж зворотного розповсюдження із зворотним зв'язком, реалізованих в середовищі MATLAB-Simulink. Були враховані чотири діючі силові трансформатори, що працюють в різних умовах оточуючого середовища та клімату, таких як: пустеля, волога, холод. Представлені практичні результати діагностики цих силових трансформаторів з використанням аналізу розчиненого газу. Практичне значення. Стисло наведено структуру та специфічні особливості старіння ізоляції обмоток силових трансформаторів та діагностики стану дефектів за допомогою штучної нейронної мережі. Далі були розроблені програми у MATLAB для автоматизації оцінки кожного методу. Ця стаття представляє ще один засіб для аналізу результатів, отриманих за допомогою програмного забезпечення delta X, що широко використовується електричною компанією в Алжирі.* Бібл. 29, табл. 15, рис. 9.
*Ключові слова:* **аналіз розчинених газів у маслі, діагностика силових трансформаторів, нейронні мережі зі зворотним зв'язком, метод Роджерса, метод МЕК.**

**Introduction.** The power transformer is a capital device in the power electrical system and we can't give up it, many researchers interested to diagnosis and protect the power transformers to improve their lifespan, their performance and their reliability [1-5]. For that, improved techniques for power transformer diagnosis and early fault detection are really important. The transformer is subject to electrical and thermal stresses. These two stresses could break down the insulating materials and release gaseous decomposition products.

According to literature [4, 6-11], the main causes of power transformer faults related gases are: overheating, corona, cellulose degradation by overheating (OH) and arcing (ARC).

Principally, gases generated from several well-known faults involved inside the active parts are represented in the Table 1 [3].

Dissolved gas analysis (DGA) is widely used for diagnosing the developing faults in power transformers,

this method is classified among effective methods. Several diagnostic criteria have been developed to interpret the dissolved gases inside the power transformers [3], the interested reader is referred to IEEE Guide [12] for more details. These methods should be able to get the relations between the dissolved gases inside a power transformer and distinguishing the type of fault that occurred. Some of these relations are clear, others are less obvious and ambiguous, or even hidden relations [13]. Yet, much of the power transformer diagnostic data requires expert hands to properly analyze, approve and interpret its results [3, 13].

Table 1
Gases generated inside the power transformer due to several faults

| Chemical symbol | Name gas |
|---|---|
| $H_2$ | Hydrogen |
| $CH_4$ | Methane |
| $C_2H_2$ | Acetylene |
| $C_2H_6$ | Ethane |
| $C_2H_4$ | Ethylene |
| CO | Carbon monoxide |
| $CO_2$ | Carbon dioxide |

Non-conventional methods are usually computer-aided, will be able to detect reliably and efficiency the incipient-faults for the inexperienced engineer. More than that, in certain cases, even the well experienced engineers can benefit further insights [13], several researches have been published in [14].

Artificial intelligence and/or optimization algorithm-based expert systems [4, 7-11, 15-18] have been developed to expose some of the secret relations to well interpret the dissolved gases in purpose of power transformer fault diagnosis.

The artificial neural network (ANN) method was also applied to this type of study, by exploiting their capacities, their properties in terms of learning and their processing of complex data in order to extract the ambiguous relations between dissolved gases inside the power transformer and the type of faults. Zhang et al. in [11] use a two-step ANN approach in order to detect faults in the diagnosis of power transformers by DGA without cellulose involved or even with cellulose involved.

Bondarenko et al. in [4] combine two powerful technique and propose a fuzzy-ANN to power transformer oil gases analysis.

Enriquez et al. in [19] propose as tools for power transformer diagnosis a K-Nearest Neighbors algorithm (K-NN) classifier with weighted classification distance, applied to a DGA data, where K-NN is one of the most fundamental and basic classifiers widely used in pattern recognition applications, this classifier was developed by T.M Cover et al in 1967 [20].

A two types of feed-forward neural network classifiers for power transformer diagnosis has been designed and used by Seifeddine et al. in [21], the two types are MLP and the Radial Basis Function (RBF). In order to train the two NN classifier the authors get the experimental data from the Tunisian Company of Electricity and Gas (STEG).

However, the diagnosis presented by [21] is not precise enough in certain types of faults such as partial discharge (PD) and temperature overheating faults. Where in PD fault condition for [21] it is only one type of fault, whereas in our work here, we separate between PD faults with low energy density and with high energy density faults.

Likewise, with regard to temperature overheating, [21] present three ambiguous overheating levels: low, medium and high, while in our study the diagnosis is precise and divided into four different states limited by clear numerical values; which more clearly show the thermal state of power transformer.

This diagnosis allows us to judge whether or not the maintenance of the power transformer is necessary. In the worst cases, immediate shutdown. This judgement is depending on the maximum temperature supported by the insulation.

**The goal of the work** is to improve the quality of diagnostics of electrical power transformer by artificial neural network tools based on two conventional methods in the case of a functional power transformer at Sétif province in east north Algeria. The type of the used ANN is Feedforward Neural Network (FNN) trained by Levenberg-Marquardt Backpropagation (LMBP). The training patterns set used to learn ANN are a practical result obtained in this functional power transformer.

**Dissolved gas-in-oil analysis.** In power transformer faults detection and analysis, dissolved gas in oil analysis (DGA) is a famous standard practice. The origin of this method goes back to 1973, where Halstead carried out of study in formation of gaseous hydrocarbons in faulty transformers based on thermodynamic assessment [11]. Under extreme thermal and electrical pressures applied to the power transformer, also under the effect of aging, mineral oils and cellulosic materials used for winding electrical insulations of a power transformer are degrading. This degradation of the material results in several types of gas emitted inside that we can use them as identifying indices of the type and intensity of the stresses. To estimate the transformer health-state, concentrations of dissolved gas in oil, relative gases proportions and gas creation rates, are analyzed and applied [13]. The most common gases used for diagnostics are given in Table 1.

In the goal to obtain each gas concentration separately we collect it from high-vacuum, then we use the gas chromatography techniques to analyze them [22]. It is possible to diagnose existing defects in the power transformers by interpretation of the gas contents. The specialist literature presents several methods for interpreting these diagnosis results [3]. Several method of power transformers diagnostic by DGA has been developed, we can classify them by two main types, classical and hybrid methods; where the classical types generally are: Key gases, Doernenburg ratio, Duval triangle, Rogers ratio, pentagon and IEC methods [6, 15, 17, 22, 23]. However, the second type generally associate one or more of the classical methods with an artificial intelligent [4, 7-11], probabilistic approach [15, 16] or with optimization method [17, 18].

**Key gas method**. This method uses four individual gas levels ($C_2H_4$, CO, $H_2$, $C_2H_2$), they are often called «key gases» to detect four types of faults; this method depends on the amount of fault gas emitted in the power transformer by the insulating oil under the effect of the varying temperatures at the chemical structure breaks [12, 13, 22]. Unfortunately, this methodology suffers from its findings or it may be incorrect or inconclusive; it can reach 50 % if it is automatically implemented with software, this error can be reduced to 30 % if it is applied manually by experienced DGA users [12].

**Ratios methods.** There are several types of ratios method [3, 12, 22] where each method assigns some combination of codes individually to a particular form of fault [22]. These methods are based on comparing the current gas ratios to the preset ratio intervals.

The principle of fault detection is simple and is as follows once a combination of codes matches the code pattern of the fault the fault is detected [13, 22]. As a result of the amount of doable code combination is larger than the amount of fault sorts, the typical result of Ratio' strategies such as Doernenburg ratios, Rogers ratios and IEC ratios is generally «no decision» [13], this is one of their important limitations, several of their limitations are described in [12], these limitations either no decision or false detection were verified in our study as shown at the end of this article in Table 15.

Other details like the variability of dissolved-gas data, loading and environmental conditions effects on these data are typically often taken into account in a real power transformer diagnostic process [24].

ANN can identify the hidden relations between the types of faults and dissolved gases in the power transformer [3] therefore, the ANN approach can be applied more carefully to solve this issue.

**Gas ratio methods.** The major benefits of using Ratios techniques in monitoring the health of the power transformer are that the problem of oil volume does not arise. Because this technique needs to compute the Gases Ratios to detect a fault, and its independent of its absolute values [3]. The Ratios techniques applied in this research are the Rogers Ratio and the IEC techniques.

**Rogers ratio method.** Four ratios are applied in this assessment: methane/hydrogen, ethane/methane, ethylene/ethane, and acetylene/ethylene. The fault diagnosis process of power transformer is achieved through a simple coding scheme based on ratio ranges. Four states of power transformer condition can be identified i.e. normal ageing, PD with or without tracking, thermal fault and electrical fault of various degrees of severity, the Table 2 give the Roger's ratio codes [3, 25]

Table 2
Roger's ratio codes

| Ratios | Range | Code |
|---|---|---|
| Methane / Hydrogen | $\leq 0.1$ | 5 |
| | $0.1 < ... < 1$ | 0 |
| | $1 \leq ... < 3$ | 1 |
| | $3 \leq$ | 2 |
| Ethane / Methane | $< 1$ | 0 |
| | $1 \leq$ | 1 |
| Ethylene / Ethane | $< 1$ | 0 |
| | $1 \leq ... < 3$ | 1 |
| | $3 \leq$ | 2 |
| Acetylene / Ethylene | $< 0.5$ | 0 |
| | $0.5 \leq ... < 3$ | 1 |
| | $3 \leq$ | 2 |

Table 3 gives the Roger's fault diagnosis table [25].

Table 3
Roger's fault diagnosis table

| N | Code | | | | Diagnosis |
|---|---|---|---|---|---|
| 1 | 0 | 0 | 0 | 0 | Normal (N) |
| 2 | 5 | 0 | 0 | 0 | Partial discharge of low energy |
| 3 | 1.2 | 0 | 0 | 0 | Overheating < 150 °C |
| 4 | 1.2 | 1 | 0 | 0 | Overheating 150 -200 °C |
| 5 | 5 | 1 | 0 | 0 | Overheating 200-300 °C |
| 6 | 0 | 0 | 1 | 0 | Conductor Overheating |
| 7 | 1 | 0 | 1 | 0 | Overheating by winding circulating current |
| 8 | 1 | 0 | 2 | 0 | Overheating by core and tank circulating current |
| 9 | 0 | 0 | 0 | 1 | Arcing of low energy |
| 10 | 0 | 0 | 1.2 | 1.2 | Arcing of high energy |
| 11 | 0 | 0 | 2 | 2 | Continuous sparking to floating potential |
| 12 | 5 | 0 | 0 | 1.2 | Partial discharge with high energy |

**IEC method.** The exclusion for the Ethane/Methane ratio was dropped from the diagnostic instructions suggested by IEC, as is indicate only a small decomposition temperature range, the IEC method is derived from the Rogers' technique. In the IEC process, three gas ratios are computed and we use it for the failure's interpretation, IEC can identify four types of faults as fellow: normal ageing, PD of low and high energy density, thermal faults and electrical faults of various degrees of severity [3, 26].

The codes for various gas ratios and their description for the IEC method are presented in Table 4 and 5. On the other hand, the downside to these Ratios approaches is that all data ranges are not covered and ratios fall beyond the reach of the tables very frequently [13]. In this work, an ANN technique was used to resolve the above limited ratios process limitation.

Table 4
IEC ratio codes

| Defined range of the gas ratio | Codes of different gas ratio | | |
|---|---|---|---|
| | $C_2H_2/C_2H_4$ | $CH_4/H_2$ | $C_2H_4/C_2H_2$ |
| < 0.1 | 0 | 1 | 0 |
| 0.1-1 | 1 | 0 | 0 |
| 1-3 | 1 | 2 | 1 |
| > 3 | 2 | 2 | 2 |

Table 5
IEC fault diagnosis table

| N | Fault type | code | | |
|---|---|---|---|---|
| 1 | No Fault | 0 | 0 | 0 |
| 2 | Partial discharge with low energy density | 0 | 1 | 0 |
| 3 | Partial discharge with high energy density | 1 | 1 | 0 |
| 4 | discharge of low energy | 1.2 | 0 | 1.2 |
| 5 | discharge of high energy | 1 | 0 | 2 |
| 6 | Overheating $T <$ 150 °C | 0 | 0 | 1 |
| 7 | Overheating 150< $T <$ 300 °C | 0 | 2 | 1 |
| 8 | Overheating 300≤ $T ≤$ 700 °C | 0 | 2 | 1 |
| 9 | Overheating ≥ 700 °C | 0 | 2 | 2 |

**DGA practical diagnostic results for different region in Algeria.** In this section we present a DGA practical diagnostic results in four different regions.

**Case 1. Mobile station 220 kV, 40 MVA of El-Meghier (province has a hot desert climate, with very little precipitation).** Through a periodic inspection,

a DGA pattern has revealed the existence of partial discharges symptoms from the critical amount extracted of $H_2$ and $C_2H_2$ (Table 6).

Table 6
DGA values of El-Meghier transformer

| Gases (PPM) | Before treatment 17 April 2001 | After treatment | |
|---|---|---|---|
| | | 6 May 2003 | 24 May 2005 |
| $H_2$ | 111 | 27 | 41 |
| $CO_2$ | 2188 | 1757 | 2737 |
| $CO$ | 293 | 316 | 419 |
| $CH_4$ | 26 | 1 | < 1 |
| $C_2H_4$ | 31 | 20 | < 1 |
| $C_2H_6$ | 9 | 14 | < 1 |
| $C_2H_2$ | 65 | < 1 | < 1 |

The advised maintenance action was then a physical treatment of the transformer via the oil purification. Several samplings were done after this operation and there was not any sign of the previous defect. Because the transformer was fitted by a forced oil flow system, we have thought about the existence of particles attracted on the surface of windings or in any region with high electric or magnetic field. These particles were removed by circulating the oil during the purification as shown in Fig. 1.

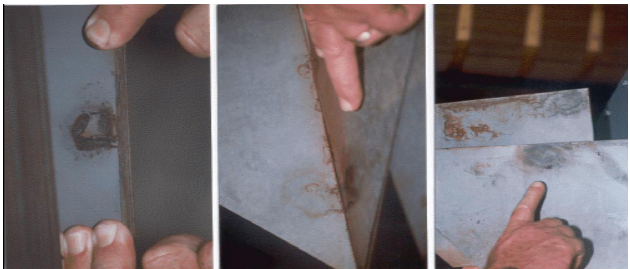

Fig. 1. Remove the particles by circulating the oil during the purification

The DGA practical measure results of El-Meghier transformer for three different years presented in Table 6 are graphically shows by Fig. 2.

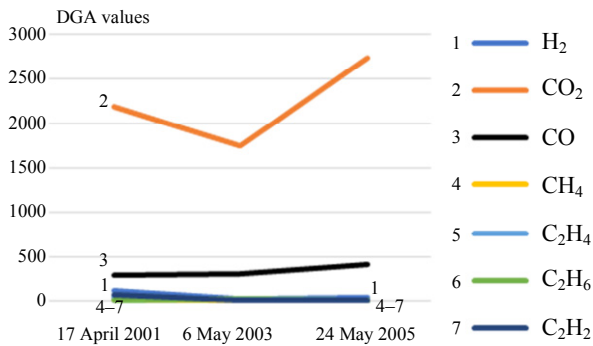

Fig. 2. DGA trends of El-Meghier transformer

**Case 2. Mobile station 220 kV 40 MVA of Sidi-Aiche (town in middle northern Algeria with a moderate and humid climate).** This transformer presented a high value of a dielectric losses factor of the oil (0.21) even though the water concentration was normal (11 ppm). The DGA revealed abnormal proportions of methane, ethane, CO and $CO_2$ concentrations giving an idea about the loading of the transformer. Furanic compounds analysis confirmed the existence of a 2-FAL (Table 7) in dangerous amount produced as result of an advanced aging process of the solid insulation. The elapsed life of the transformer was estimated at 80 %. It was then very necessary to check the reliability of our protections and to rate the transformer at a moderate loading until its inspection.

Table 7
DGA values of Sidi-Aiche transformer

| Gases (ppm) | Values | Dielectric parameters | Values |
|---|---|---|---|
| $H_2$ | 11 | $H_2O$ in ppm | 11 |
| $CO_2$ | 1944 | | |
| $CO$ | 597 | tg$\delta$ | 0.21 |
| $CH_4$ | 101 | | |
| $C_2H_4$ | < 1 | | |
| $C_2H_6$ | 110 | 2-Furfuraldehyd in PPM | 5.79 |
| $C_2H_2$ | < 1 | | |

**Case 3. Power transformer 60 kV of Akbou (industrial town in northern Algeria with a moderate and humid climate).** An excess of combustible gases concentration was registered through a DGA as shown in Table 8; a more dangerous one was the acetylene. A gases ratio confirmed a thermal overheating involving a solid insulation. Perhaps it touched the current flow system as LTC contacts, leads contacts, etc. This is the most critical case that may result in catastrophic failure. The main action was to stop the transformer for an internal inspection.

Table 8
Gases concentration of Akbou unit

| Gases | Values (ppm) |
|---|---|
| $H_2$ | 1443 |
| $CO_2$ | 13561 |
| $CO$ | 934 |
| $CH_4$ | 3899 |
| $C_2H_4$ | 600 |
| $C_2H_6$ | 1115 |
| $C_2H_2$ | 113 |

A hot spot was found at the contacts level of one connection of the no load tap changer with apparent signs of tracking, cocking, erosion and overheating with the surrounded insulation as illustrated by Fig. 3.

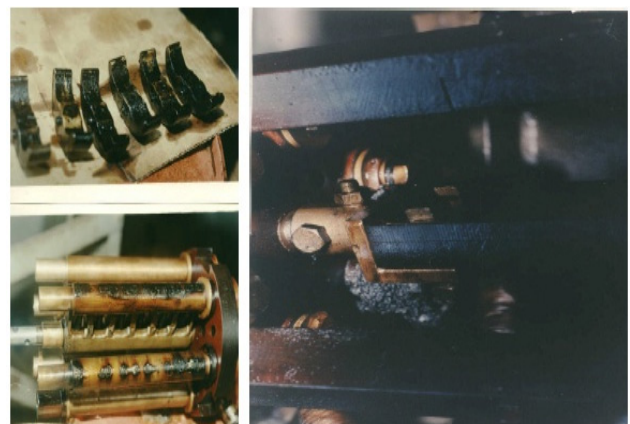

Fig. 3. Remove the particles by circulating the oil during the purification

**Case 4. Autotransformer 220/150 kV of Darguina (town in northern Algeria with cold and very humid climate).** This unit exhibited a high arcing in the oil without involving the solid insulation. It has been treated several times but there was no improvement of the situation because successive Buckholz alarms have been registered after (Table 9). The power transformer was de-energized and submitted to further analysis and diagnostic tests.

Table 9
DGA history of Darguinas autotransformer

| Gas (ppm) | 17 April 2001 | 14 March 2003 | 23 May 2005 |
|---|---|---|---|
| $H_2$ | 107 | <1 | 645 |
| $CO_2$ | 1414 | 434 | 2099 |
| CO | – | 40 | 217 |
| $CH_4$ | 27 | <1 | 45 |
| $C_2H_4$ | 25 | <1 | 51 |
| $C_2H_6$ | 18 | <1 | <1 |
| $C_2H_2$ | 65 | 7 | 326 |

Complementary electrical tests have been investigated in order to assess the integrity of the current circuit and the windings condition. The results were normal (Table 10 and Table 11) and the problem causing this generation of gases still not detectable.
- Insulation measure: Primary/Ground = 7 GΩ;
- Secondary/Ground = 5 GΩ.

Table 10
Winding resistance measurement with different tap position

| | 150 kV | | | 220 kV | | |
|---|---|---|---|---|---|---|
| Position | A/N | B/N | C/N | A/N | B/N | C/N |
| 5 | 3.53 | 3.57 | 3.53 | 5.02 | 5.03 | 5.01 |
| 4 | 3.48 | 3.51 | 3.49 | 4.83 | 4.92 | 4.83 |
| 3 | 3.49 | 3.50 | 3.48 | 4.66 | 4.68 | 4.61 |
| 2 | 3.49 | 3.50 | 3..50 | 4.79 | 4.78 | 4.77 |
| 1 | 3.50 | 3.50 | 3.50 | 4.92 | 5.02 | 4.95 |

Table 11
Transformation ration

| Position | AB/ab | AC/ac | BC/bc | Nameplate value |
|---|---|---|---|---|
| 1 | 1.553 | 1.557 | 1.555 | 1.557 |
| 2 | 1.502 | 1.504 | 1.504 | 1.506 |
| 3 | 1.452 | 1.453 | 1.453 | 1.455 |
| 4 | 1.401 | 1.403 | 1.402 | 1.404 |
| 5 | 1.350 | 1.352 | 1.351 | 1.353 |

This mode of failure may be also created by the magnetic stray flux. The internal inspection revealed an overheating localized in the connection bolts of the core yoke caused by arcing of a formed closed loop. Fig. 4 illustrates a hotspot localized in power transformer.

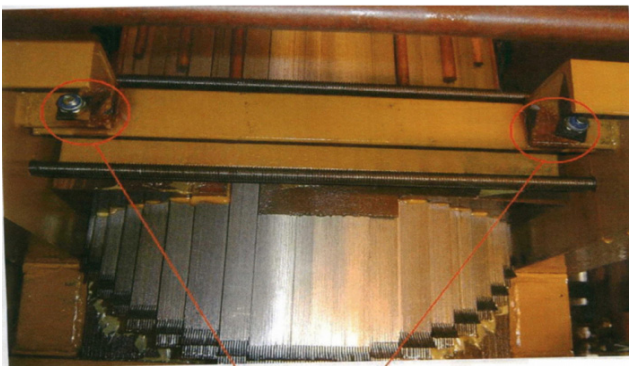

Fig. 4. Hotspot localized

**Artificial Neural Networks.** ANN has demonstrated their capacity in a varied engineering application such as, estimation, process control, diagnostics [27], it is ANN's ability and malleability to approximate functions, data meaning and classifications that have nominated it to be a proposed and promising solution that can be used in various types of complex issues. These properties are particularly very significant when the process or the state variables of the process model are nonlinear, poorly identified and uncertain, therefore hard to model by known traditional methods.

An ANN is a complex and dynamic system of different topology which is composed by weights linked together either complete or partial depending on the type of ANN, these weights are an element of complex data processing. We can summarize the principal advantages when we apply ANN in the diagnosis process of power transformer by dissolved gas as follow [28]:
- high learning capacity hence generalization of the developed tool;
- development of electronic circuits make the hardware implementation easier;
- a great capability for model a complex system based on their inputs and outputs without needing to know exactly the mathematical model;
- minimize the time required to give the final diagnosis results by eliminating the time required to conduct laboratory experiments

Several types of ANN are presented in the literature in terms of topology, backpropagations algorithm and learning mode, among these types Multi-Layer Perceptron (MLP) is the most widely used [27]. Fig 5 shows the structure of MLP, which is a type of ANN used in this research. The weight outputs are computed as follows [7]

$$S_j = f\left(\sum_{i=1}^{n} X_i W_{ij} + \theta_j\right), \quad (1)$$

where $X_i$ are the network inputs will be described in the following section; $W_{ij}$ translate the weight-connection between the input neuron $i$ and the neighboring hidden neuron $j$; $\theta_j$ is the bias of the $j$-th hidden neuron; $f$ is the transfer function or also called activation function.

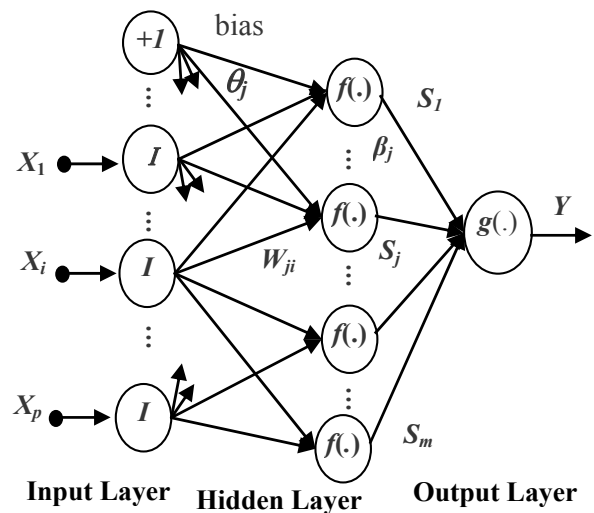

Fig. 5. ANN topology

**Topology of ANN used.** In this application, a MLP type ANNs with supervised training based on Levenberg-Marquardt Backpropagation (LMBP) algorithm is used. LMBP which remains until now the most widely applied. According to the literature related to this topic, LMBP has a good robustness, known for its high efficiency and relatively fast require much less iteration to converge compared to some methods, for this many researchers suggest to use it for power transformer diagnosis based on DGA.

The optimal size of the ANN that produces the best results is one of the most frequently phrased questions in the ANN computation framework. Even though numerous «hints and tips» such as suggestions have been highlighted so far by many researchers, but there is still no straightforward answer to this ANN issue [29].

We using Matlab/Simulink environment to design this ANN, we notice that each method has its self inputs and should be match the same as the number of neurons in the input layer, the same thing for the output layer the number of units should be the same outputs for each method.

In this application case, five main preliminary gases for a failure in power transformers: Acetylene ($C_2H_2$), Ethane ($C_2H_6$), Ethylene ($C_2H_4$), Hydrogen ($H_2$) and methane ($CH_4$) are elected as the inputs characteristics as illustrated by Fig. 6.

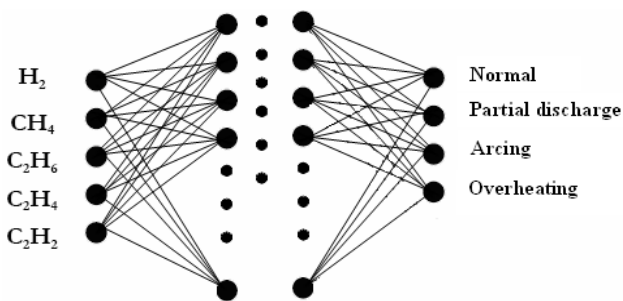

Fig. 6. Input and output of the neural network

**Input layer.** The ANN inputs are a vector contains the Rogers and IEC Ratios. Generally, the number of neurons is equal to the number of inputs and the activation function are linear. Each method has its inputs of the follow gases: Acetylene ($C_2H_2$), Ethane ($C_2H_6$), Hydrogen ($H_2$), Ethylene ($C_2H_4$) and Methane ($CH_4$) in ppm.

**Output layer.** Contains the desired outputs of the artificial neural network, in our case are the type of faults if exist or the power transformer is healthy as presented by Fig. 6. The outputs are: discharge of low energy and high energy, arcing, overheating, or normal case which means healthy state. Each method has a specific fault detection, the activate function generally are linear function.

**Hidden layer.** The wide majority of studies in classification problems use a single layer or at most two hidden layers. There is no law to determine exactly the number of neurons in the hidden layer and it is defined by trial in order to get the minimum error.

**Transfer function.** At the hidden layer, the activation function most widely used in the networks of neurons is sigmoid tangent (logsig), Fig. 7 show its curvature.

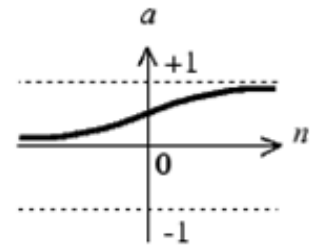

$a = \text{logsig}(n)$
Fig. 7. Activation function

The optimized ANNs used in the process of power transformer diagnosis is carried out by adjusting the number of hidden layer and the number of units in each hidden layer of the MLP. Rogers Ratios and IEC Ratios methods are two Ratios approaches investigated in this study. We are using the cross-validation error method over multiple sets both for Rogers and IEC training data to find the best results.

**Input data.** The inputs data are the gas concentrations given by the chromatographic analysis tests of the oil samples, these samples are taken over the life of the transformer (training step), or suspected in the data set sample (net value under regular conditions of use). Table 12 displays the sample data set in concentrations of principal gases in ppm.

Table 12
Samples data

| Samples | $H_2$ | $CH_4$ | CO | $CO_2$ | $C_2H_4$ | $C_2H_6$ | $C_2H_2$ | Known fault |
|---|---|---|---|---|---|---|---|---|
| 1 | 17 | 15 | 292 | 6956 | 78 | 20 | 35 | ARC |
| 2 | 1046 | 2809 | 681 | 7820 | 321 | 675 | 7 | PD |
| 3 | 127 | 76 | 879 | 3471 | 23 | 32 | 49 | ARC |
| 4 | 11 | 101 | 597 | 1944 | 110 | <1 | <1 | OH |
| 5 | 107 | 27 | – | 1414 | 18 | 25 | 65 | ARC |
| 6 | 39 | 33 | 991 | 3280 | 9 | 7 | 2 | Normal |
| 7 | 72 | 278 | 53 | 610 | 176 | 289 | <1 | OH |
| 8 | 1 | 39 | 361 | 4081 | 9 | 36 | 1 | Normal |
| 9 | 111 | 26 | 293 | 2188 | 31 | 9 | 65 | PD |
| 10 | 1443 | 3899 | 934 | 13561 | 600 | 1115 | 113 | OH |

The database of diagnostics was created at the power transformer park in the Sétif region, then has helped us to make sure for the sensitivity of our programs and the degree of its reproducibility; where several comparisons of the acquired results was making with the DELTA X software used in the Sonelgaz-GRTE laboratories or the shared data and others are considered to view the established convergences.

**Application of ANN to DGA diagnosis using IEC method.** The input vector

$$I = [I_1, I_2, I_3] = \left[ code\left(\frac{Acetylene(C_2H_2)}{Ethylene(C_2H_4)}\right), \right.$$
$$\left. code\left(\frac{Ethylene(C_2H_4)}{Hydrogen(H_2)}\right), code\left(\frac{Ethylene(C_2H_4)}{Ethane(C_2H_6)}\right) \right],$$

these codes are extracted following the IEC method shown in Table 4.

The output vector *O* in IEC method has nine detectable defects as shown in Table 13, $O = [O_1, ..., O_9]$.

Table 13
Database trained by IEC Ratios Method

| Inputs | | | Outputs | | | | | | | | |
|---|---|---|---|---|---|---|---|---|---|---|---|
| $I_1$ | $I_2$ | $I_3$ | $O_1$ | $O_2$ | $O_3$ | $O_4$ | $O_5$ | $O_6$ | $O_7$ | $O_8$ | $O_9$ |
| 0 | 0 | 0 | 1 | 0 | 0 | 0 | 0 | 0 | 0 | 0 | 0 |
| 0 | 1 | 0 | 0 | 1 | 0 | 0 | 0 | 0 | 0 | 0 | 0 |
| 1 | 1 | 0 | 0 | 0 | 1 | 0 | 0 | 0 | 0 | 0 | 0 |
| 1,2 | 0 | 1,2 | 0 | 0 | 0 | 1 | 0 | 0 | 0 | 0 | 0 |
| 1 | 0 | 2 | 0 | 0 | 0 | 0 | 1 | 0 | 0 | 0 | 0 |
| 0 | 0 | 1 | 0 | 0 | 0 | 0 | 0 | 1 | 0 | 0 | 0 |
| 0 | 2 | 0 | 0 | 0 | 0 | 0 | 0 | 0 | 1 | 0 | 0 |
| 0 | 2 | 1 | 0 | 0 | 0 | 0 | 0 | 0 | 0 | 1 | 0 |
| 0 | 2 | 2 | 0 | 0 | 0 | 0 | 0 | 0 | 0 | 0 | 1 |

The neural network with the IEC method is composed by three unites in input layer, the hidden layers comprise a variable number of unites and the output layer has nine unites. The unites of the output layers generate a real number between 0 and 1 indicating the possibility of presence of a fault among the 9 faults designated by IEC instructions norms. The models of formation for the IEC technique are presented in Table 13.

**Application of ANN to DGA diagnosis using Rogers method.** The vector of inputs

$$I = [I_1, I_2, I_3, I_4] = \left[\left(\frac{Methane}{Hydrogen}\right), \left(\frac{Ethane}{Methane}\right), \left(\frac{Ethylene}{Ethane}\right), \left(\frac{Acetylene}{Ethylene}\right)\right],$$

these codes are extracted following the Rogers method shown in Table 2.

The output vector *O* in Rogers method has 12 detectable defects as shown in Table 2, $O = [O_1, ..., O_{12}]$.

The neural network with the Rogers method is composed by 4 unites in the input layer; the hidden layers contain a variable number of neurons and the output layers has twelve unites. The neurons of the output layers generate a real number between 0 and 1 giving the likelihood of the presence of a fault amongst the twelve faults designated by the Rogers' procedure. The models of formation for the Rogers technique are exposed by Table 14 below.

Figure 8 shows the ANN training performance step trained by LMBP algorithm based on IEC set.

Figure 9 shows the ANN training performance step trained by LMBP algorithm based on Rogers set.

**Results and discussions.** The two artificial neural networks elaborate previously needs to be checked after the training step is done. In order to test and verify the robustness and the best performance of the trained ANN, a data set pattern of known cause of power transformer faults was considered in 10 test samples given in Table 12.

Table 14
Data sets trained by Rogers Ratios technique

| Inputs | | | | Outputs | | | | | | | | | | | |
|---|---|---|---|---|---|---|---|---|---|---|---|---|---|---|---|
| $I_1$ | $I_2$ | $I_3$ | $I_4$ | $O_1$ | $O_2$ | $O_3$ | $O_4$ | $O_5$ | $O_6$ | $O_7$ | $O_8$ | $O_9$ | $O_{10}$ | $O_{11}$ | $O_{12}$ |
| 0 | 0 | 0 | 0 | 1 | 0 | 0 | 0 | 0 | 0 | 0 | 0 | 0 | 0 | 0 | 0 |
| 5 | 0 | 0 | 0 | 0 | 1 | 0 | 0 | 0 | 0 | 0 | 0 | 0 | 0 | 0 | 0 |
| 1,2 | 0 | 0 | 0 | 0 | 0 | 1 | 0 | 0 | 0 | 0 | 0 | 0 | 0 | 0 | 0 |
| 1,2 | 1 | 0 | 0 | 0 | 0 | 0 | 1 | 0 | 0 | 0 | 0 | 0 | 0 | 0 | 0 |
| 0 | 1 | 0 | 0 | 0 | 0 | 0 | 0 | 1 | 0 | 0 | 0 | 0 | 0 | 0 | 0 |
| 0 | 0 | 1 | 0 | 0 | 0 | 0 | 0 | 0 | 1 | 0 | 0 | 0 | 0 | 0 | 0 |
| 1 | 0 | 1 | 0 | 0 | 0 | 0 | 0 | 0 | 0 | 1 | 0 | 0 | 0 | 0 | 0 |
| 1 | 0 | 2 | 0 | 0 | 0 | 0 | 0 | 0 | 0 | 0 | 1 | 0 | 0 | 0 | 0 |
| 0 | 0 | 0 | 1 | 0 | 0 | 0 | 0 | 0 | 0 | 0 | 0 | 1 | 0 | 0 | 0 |
| 0 | 0 | 1,2 | 1,2 | 0 | 0 | 0 | 0 | 0 | 0 | 0 | 0 | 0 | 1 | 0 | 0 |
| 0 | 0 | 2 | 2 | 0 | 0 | 0 | 0 | 0 | 0 | 0 | 0 | 0 | 0 | 1 | 0 |
| 5 | 0 | 0 | 1,2 | 0 | 0 | 0 | 0 | 0 | 0 | 0 | 0 | 0 | 0 | 0 | 1 |

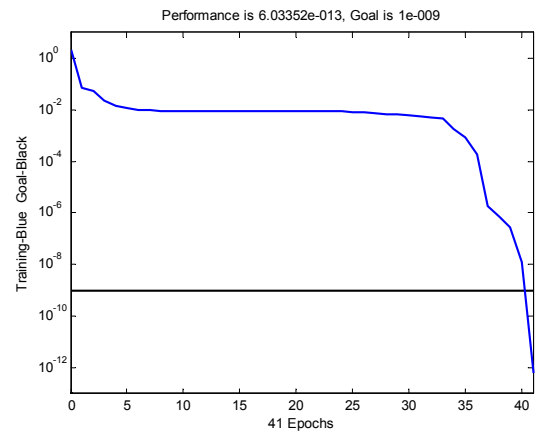

Fig. 8. Training performance using LMBP with IEC Ratios method

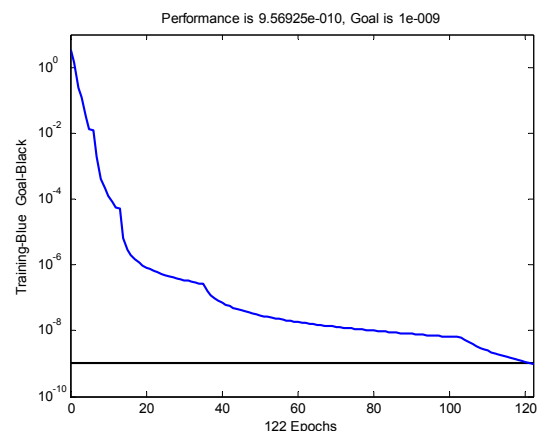

Fig. 9. Training phase performance using LMBP and Rogers Ratios method

Table 15 summarizes the findings obtained by the two ANN during testing. According to these results, ANN associates to Rogers Ratios methods was correctly diagnosed 8 faults out of 10 faults. However, ANN associated with IEC Ratios methods only detect 7 faults.

According to our results, the ANN solves all limitations of type no decision of the traditional methods

(IEC and Roger) and give a decision, for ANN based Roger ration method all these decisions are correct compared the experiment. However, for ANN based on IEC method the decision is wrong for the sample 2 and 8. In other hand, the improved IEC by ANN correct the diagnosis result of traditional IEC for the sample 1. As illustrated by Table 15 all traditional or ANN-based methods give wrong results for samples 2 and 8, and the work should be done by more efficient and robust methods as a deep learning.

Table 15
Comparison results between ANN DGA and traditional DGA based on Rogers and IEC

| Samples | Actual fault | Fault from traditional IEC method | Fault from traditional Rogers method | Predicted fault from ANN | |
|---|---|---|---|---|---|
| | | | | IEC Method | Rogers Method |
| 1 | ARC | PD | No decision | ARC | ARC |
| 2 | PD | No decision | OH | OH | OH |
| 3 | ARC | No decision | ARC | ARC | ARC |
| 4 | OH | OH | No decision | OH | OH |
| 5 | ARC | No decision | No decision | ARC | ARC |
| 6 | N | PD | OH | PD | Normal |
| 7 | OH | No decision | OH | OH | OH |
| 8 | N | No decision | OH | OH | OH |
| 9 | PD | PD | ARC | PD | OH |
| 10 | OH | No decision | OH | OH | OH |

**Conclusions.**

The dissolved gas analysis has been acknowledged as an important instrument in the health-state monitoring and the diagnosis of faults of power transformer. The principal benefit of using ratios techniques is that only ratios of gases are needed in the computation process, therefore, the oil quantity that is participated in gas dissolution is not needed. On the other hand, the downside is that they struggle to cover all dissolved gas analysis data set ranges. Each traditional method has their own limitations generally no decision or false decision as found in this work.

Artificial neural network is proposed as a solution and is then used to solve this inconvenience and treat cases not identified by classical techniques in order to an almost reliable diagnosis. The majority of limitations has been removed and the diagnostic results has been improved, the findings obtained through artificial neural network are extremely reliable compared to the traditional methods, where for IEC method the accuracy has been increased for 20 % to 70 % and from 40 % to 70 % for Roger method. However, in some sample, all methods either based artificial neural network or traditional are misleading and give a false diagnosis, so, the health of power transformer is vulnerable. Finally, as conclusion this improved results need more performed by more effective method such as deep learning.

**Conflict of interest.** The authors declare that they have no conflicts of interest.

*Lahcene Bouchaoui* [1], *PhD Student*,
*Kamel Eddine Hemsas* [1], *Full Professor*,
*Hacene Mellah*[2], *PhD*, *Associate Professor*,
*Saadeddine Benlahneche* [3], *Chief of Electricity Transmission*,
[1] Electrical Engineering Department,
University Ferhat Abbas Setif 1,
Setif, Algeria,
e-mail: bouchaoui.lahcene@gmail.com,
hemsas_ke_dz@univ-setif.dz,
[2] Electrical Engineering Department,
Bouira University,
Bouira, Algeria,
e-mail: has.mel@gmail.com (Corresponding author),
[3] GRTE /SONELGAZ,
Setif, Algeria,
e-mail: benlahneche.saadeddine@grte.dz